\documentclass{appolb}
\usepackage{graphicx}
\usepackage{cite} 
\usepackage{hyperref}

\begin{document}
\title{My friends and my path in physics}
\author{Mirzayusuf Musakhanov
\address{Institute of Theoretical Physics, National University of Uzbekistan}
\\[3mm]
\address{Tashkent 100174, Uzbekistan}}

\maketitle
\begin{abstract}
Dedication to my untimely departed friends, Dmitry Igorevich Diakonov, Viktor Yur'evich Petrov and Maxim Vladimirovich Polyakov. 
\end{abstract}
  
\section{Introduction}
I have received an offer from Igor Strakovsky to participate in a special volume of Acta Physica Polonica B dedicated to the outstanding theoretical physicists from Leningrad (now St. Petersburg) Dmitry Igorevich Diakonov (Mitya), Viktor Yur'evich Petrov (Vitya) and Maxim Vladimirovich Polyakov, who sadly passed away and no longer with us.

I was lucky to get to know them when Mitya and Vitya began to develop intensively the ideas of QCD instanton vacuum theory. It was the time of 1980-88.  At that time they had some students, whom I met later, Maxim Polyakov and Pavel Pobylitsa.

\section{The time of 1980-88}
At that time, I was working on the quark structure of hadrons and it was important for me to discuss my ideas (for example, the chiral bag model~\cite{Musakhanov:1980qs,Musakhanov:1980va,Musakhanov:1981ec,Musakhanov:1981ra,Musakhanov:1981rg,Israilov:1981wf,Musakhanov:1982gh,Israilov:1982as,Musakhanov:1984qy,Musakhanov:1984hk,Musakhanov:1985wz,Musakhanov:1987xe,Dorokhov:1988sk,Musakhanov:1991dw}) at seminars in ITEP (Moscow), JINR (Dubna), and, of course, in LINP (Leningrad), where I met Mitya and Vitya at LINP school on high energy physics. I realized that the topic of vacuum in gauge field theories is very relevant and we held special conferences in 1985 and 1987 on this topic in Tashkent, where the most outstanding Soviet experts on quantum field theory of that time, including Mitya and Vitya, were present. Their papers on QCD  instanton vacuum theory of that time were a real bomb\cite{Diakonov:1983dt,Diakonov:1983hh,Diakonov:1984xj,Diakonov:1984vw,Diakonov:1984vw,Diakonov:1984trb,Diakonov:1985eg,Diakonov:1987ty,Diakonov:1988ju}. I realized that they are very close to the solution of the problem about quark-gluon structure of hadrons, which, of course, are just excitations of the QCD vacuum.  

\section{The time of 1990-2000}

It was important to find a convincing proof of the correctness of Mitya's and Vitya's ideas about the QCD vacuum, which was done in our papers on checking the fulfillment of the axial anomaly theorems in their theory~\cite{Musakhanov:1996cv,Musakhanov:1996qf,Musakhanov:2002xa}.  By that time the Ruhr University (Bochum, Germany) became the Mecca of hadron physics because of works done by Mitya, Vitya, Maxim and Pavel. To discuss my results on the development and application of the instanton vacuum to hadron 
physics~\cite{Musakhanov:1998wp,Musakhanov:1999zt,Musakhanov:2001pc,Musakhanov:2002vu,Yakhshiev:2004pj,Kim:2004hd,Kim:2005jc,Goeke:2007bj,Goeke:2007nc,Goeke:2010hm}, I had arranged a seminar in Bochum with Maxim during my stay at ICTP in Trieste. After the seminar, Bochum became a frequent destination for me. I met many interesting people there, such as Klaus Goeke, Hyun-Chul Kim and others. Later, together with Klaus and other colleagues we organized a conference in Tashkent in 2003, where core topic was a QCD instantons. 
\section{the time of 2000-present time}
In recent years, our group from the National University of Uzbekistan has been collaborating with the group of Huyn-Chul from Inha University (Korea) on the application of the instanton vacuum theory developed by Mitya and Vitya  particularly to systems of heavy quarks and systems of heavy and light quarks~\cite{Nam:2008ff,Musakhanov:2011xx,Musakhanov:2014fya,Yakhshiev:2016keg,Musakhanov:2017gym,Musakhanov:2017erp,Musakhanov:2018sdu,Musakhanov:2018gho,Musakhanov:2019ago,Musakhanov:2020hvk,Musakhanov:2021gof,Musakhanov:2021rex,Musakhanov:2023dsn}. The results show that the role of interactions induced by instantons is crucial in the physics of systems consisting of heavy and light quarks~\cite{Hong:2024ptu}. 
 
\section{Instead of conclusion} 
 
I am happy that in my life I had scientific discussions, talks and discussions of the world around us with my wonderful friends Mitya, Vitya and Maxim.
It is a sad to recognize that they left us so early. 
 

\end{document}